\documentclass[aps,prd,groupedaddress,showpacs,nofootinbib,twocolumn]{revtex4}
\usepackage{graphicx,amssymb,bm}

\newcommand{\be}{\begin{equation}}
\newcommand{\ee}{\end{equation}}
\newcommand{\bea}{\begin{eqnarray}}
\newcommand{\eea}{\end{eqnarray}}
\newcommand{\beaa}{\begin{eqnarray*}}
\newcommand{\eeaa}{\end{eqnarray*}}

\newcommand{\bmr}{\bm{r}}

\newcommand{\nn}{\nonumber \\}

\newcommand{\e}{{\rm e}}

\begin{document}

\tolerance=5000

\title{Modified $f(R)$ gravity consistent with realistic cosmology:
from matter dominated epoch to dark energy universe}
\author{Shin'ichi Nojiri}
\email{nojiri@phys.nagoya-u.ac.jp}
\affiliation{Department of Physics, Nagoya University, Nagoya 464-8602. Japan}
\author{Sergei D. Odintsov\footnote{also at Lab. Fundam. Study, Tomsk State
Pedagogical University, Tomsk}}
\email{odintsov@ieec.uab.es}
\affiliation{Instituci\`{o} Catalana de Recerca i Estudis Avan\c{c}ats (ICREA)
and Institut de Ciencies de l'Espai (IEEC-CSIC),
Campus UAB, Facultat de Ciencies, Torre C5-Par-2a pl, E-08193 Bellaterra
(Barcelona), Spain}

\begin{abstract}

We develop the  general scheme for modified $f(R)$ gravity
reconstruction from any realistic FRW cosmology. We formulate several
versions of modified gravity compatible with Solar System tests  where the
following sequence of cosmological epochs occurs:
a. matter dominated phase (with or without usual matter), transition from
decceleration to acceleration, accelerating epoch consistent with recent
WMAP data
b.$\Lambda$CDM cosmology without cosmological constant.
As a rule, such modified gravities are expressed implicitly (in terms of
special functions) with late-time asymptotics of known type (for
instance, the model with negative and positive powers of curvature).
In the alternative approach, it is demonstrated that even simple versions
of modified gravity may lead to the unification of matter dominated and
accelerated phases at the price of the introduction of compensating dark
energy.

\end{abstract}

\pacs{11.25.-w, 95.36.+x, 98.80.-k}

\maketitle

\section{Introduction}

Modified gravity is extremely promising approach to dark energy.
Within gravitational alternative for dark energy (for review, see \cite{review}),
the cosmic speed-up is explained by the  universe expansion
where some sub-dominant terms (like $1/R$\cite{c,CDTT} or $lnR$\cite{grg}
which may be caused by string/M-theory\cite{no1}) may become essential at small
curvature. It also explains naturally the unification of earlier and later
cosmological epochs as the manifestation of different role of
gravitational terms relevant at small and large curvature as it happens in
the model with negative and positive powers of curvature \cite{NO}.
Moreover, modified gravity may serve as dark matter.

Special attention is paid to $f(R)$ modified gravity which may be
constrained from cosmological/astrophysical observational data
\cite{constrain} and solar system tests \cite{newton,NO,dolgov,faraoni}.
Recently, very interesting attempt to constrain such model (with positive
and negative powers of curvature) from fifth
force/BBN considerations has been made in \cite{hall}
where it was shown that it is not easy to fulfil the known constraints
and to describe the sequence of known cosmological epochs
within the simple theory with positive and negative
powers of curvature\cite{NO}. The cosmological dynamics of $1/R$ and other related
$f(R)$ theories leading to late-time acceleration has been studied in
refs.\cite{c,CDTT,NO,grg,cosmology}
while SdS black holes solutions were discussed in \cite{sds}.

In the situation when General Relativity can not naturally describe the dark
energy epoch of the univere the search of alternative, modified gravity
which is consistent with Solar System tests/observational data is of
primary interest. Of course, it is too strong (at least, at first step) to
request that such theory
should correctly reproduce all known sequence of cosmological epochs
(including inflationary universe where quantum effects may be essential).
Nevertheless, it is reasonable to constrain such a theory (if this is
really alternative theory for General Relativity) by the condition
that it reproduces well established sequence of classical cosmological phases
(matter dominated phase, transition from deceleration to acceleration and
current universe speed-up) being consistent with Solar System tests. In
the present paper we construct several examples of such modified gravity
where $f(R)$ is presented implicitly, in terms of special functions.

The important remark is in order. It is well-known fact that arbitrary
$f(R)$ gravity may be presented in mathematically-equivalent form as
minimal scalar-tensor theory (Einstein gravity with scalar
self-interacting field). Even more, it was shown \cite{capo,CNOT}
that $f(R)$ gravity may be formulated in mathematically-equivalent form as
Einstein gravity with ideal field having inhomogeneous equation of state.
Then, it may look that it is not necessary to study modified gravity and
it is enough to limit the consideration only by Einstein gravity with
scalars and (or) ideal fluid. However, the situation is more complicated.
For instance, modified gravity of specific form which describes acceptable
accelerating universe (with realistic effective equation of state)
is not physically equivalent to scalar-tensor theory \cite{capo,CNOT}.
Hence, the equivalent scalar-tensor gravity may not lead to accelerating
FRW universe (or, it may lead but with significally different effective
equation of state).  The corresponding examples were given in \cite{capo,CNOT}. Moreover,
specific form of modified gravity leads to specific form of scalar
potential. As a result, such specific modified gravity may comply with
Solar System tests (acceptable Newton law, etc) while the corresponding
scalar-tensor theory may not comply with it and vice-versa.
Hence, one should consider all three classes of theories: modified
gravity, scalar-tensor gravity and Einstein gravity with ideal fluid for
description of dark energy  and early universe.
One should fit all these three classes of theories with astrophysical and
cosmological constraints (the corresponding cosmological parameters are
defined from different bounds in \cite{tegmark}) in order to find finally
what is the realistic gravitational theory compatible with observational data. In principle,
In this work we are not so ambitious to comply with all bounds. We intend
to present the first realistic example of modified gravity which is
compatible with Solar System tests, which is cosmologically viable
(sequence of matter dominated phase, transition from decceleration to
acceleration, and acceleration phases) and which may lead even to
$\Lambda$CDM cosmology.

 The paper is organized as follows.
In the next section we present general formulation to reconstruct the
modified $f(R)$ gravity for any FRW given cosmology (using the auxiliary
scalar field).  This formulation is applied to work out
several models. The explicit example of the model where
matter-dominated phase may be realized by pure $f(R)$ gravity (no matter)
with subsequent transition to acceleration phase is presented.
It is constructed the model where the function $f(R)$ is expressed in
terms of Gauss hypergeometric functions and where the standard
$\Lambda$CDM cosmology is reproduced.
For the specific version of the modified $f(R)$ gravity with matter it is
shown that not only matter dominated phase with subsequent transition to
acceleration occurs, but the acceleration epoch complies with three years
WMAP data. In order to ensure that transition from decceleration to
acceleration indeed occurs, the (in)stability analysis of the
cosmological solutions is fulfilled.
We also demonstrate that corrections to Newton law for suggested versions
of modified gravity are negligible. Third section is devoted to
consideration of the modified gravity model \cite{NO} with compensating
dark energy (ideal fluid). It is shown that the role of such compensating dark
 energy may be
to ensure the transition from matter dominated to acceleration phase
while during the current speed-up such compensating dark energy quickly
disappears. Some outlook is given in the Discussion section. In the
Appendix it is shown that our formulation is just equivalent to standard
metric formulation of $f(R)$ gravity (without extra scalar).

\section{Reconstruction of modified gravity which describes matter
dominated and accelerated phases}

\subsection{General formulation}

In the present section we develop the general formulation of the
reconstruction scheme for modified gravity with $f(R)$ action.
It is shown how any cosmology may define the implicit form of the function
$f$. The starting action
of modified gravity is:
\be
\label{FR1}
S=\int d^4 x \sqrt{-g}f(R)\ .
\ee
First we consider the proper Hubble rate $H$, which describes
the evolution of the universe, with
radiation dominance, matter dominance, and accelerating expansion.
It turns out that one can find $f(R)$-theory realizing such a
cosmology (with or without matter). The construction is not
explicit and it is necessary to solve the second order differential
equation and algebraic equation. It shows, however, that, at
least, in principle, we could obtain any cosmology by properly
reconstructing a function $f(R)$ on theoretical level.

The equivalent form of above action is
\be
\label{PQR1}
S=\int d^4 x \sqrt{-g} \left\{P(\phi) R + Q(\phi) + {\cal L}_{\rm matter}\right\}\ .
\ee
Here $P$ and $Q$ are proper functions of the scalar field $\phi$
and ${\cal L}_{\rm matter}$ is the matter Lagrangian density.
Since the scalar field does not have a kinetic term, it may be regarded
 as an auxiliary field (compare with ideal fluid representation of $f(R)$
gravity \cite{capo}). In fact, by the variation of $\phi$,
it follows
\be
\label{PQR2}
0=P'(\phi)R + Q'(\phi)\ ,
\ee
which may be solved with respect to $\phi$:
\be
\label{PQR3}
\phi=\phi(R)\ .
\ee
By substituting (\ref{PQR3}) into (\ref{PQR1}), one obtains $f(R)$-gravity:
\bea
\label{PQR4}
S&=&\int d^4 x \sqrt{-g} \left\{f(R) + {\cal L}_{\rm matter}\right\}\ , \nn
f(R)&\equiv& P(\phi(R)) R + Q(\phi(R))\ .
\eea

By the variation of the action (\ref{PQR1}) with respect to the
metric $g_{\mu\nu}$, we obtain
\bea
\label{PQR5}
0&=&-\frac{1}{2}g_{\mu\nu}\left\{P(\phi) R + Q(\phi) \right\}
 - R_{\mu\nu} P(\phi) \nn && + \nabla_\mu \nabla_\nu P(\phi)
 - g_{\mu\nu} \nabla^2 P(\phi) + \frac{1}{2}T_{\mu\nu}\ .
\eea
The   equations corresponding to standard spatially-flat FRW universe are
\bea
\label{PQR6}
0&=&-6 H^2 P(\phi) - Q(\phi) - 6H\frac{dP(\phi(t))}{dt} + \rho \ ,\\
\label{PQR7}
0&=&\left(4\dot H + 6H^2\right)P(\phi) + Q(\phi) \nn
&& + 2\frac{d^2 P(\phi(t))}{dt} + 4H\frac{d P(\phi(t))}{dt} + p\ .
\eea
By combining (\ref{PQR5}) and (\ref{PQR6}) and deleting
$Q(\phi)$, we find the following equation
\be
\label{PQR7b}
0=2\frac{d^2 P(\phi(t))}{dt^2} - 2 H \frac{dP(\phi(t))}{d\phi} + 4\dot H P(\phi) + p + \rho\ .
\ee
As one can redefine the scalar field $\phi$ properly, we may choose
\be
\label{PQR8}
\phi=t\ .
\ee
It is assumed that $\rho$ and $p$ are the sum from the contribution of the
matters
with a constant equation of state parameters $w_i$.
Especially, when it is assumed a combination of the radiation and dust,
one gets the standard expression
\be
\label{PQR9}
\rho=\rho_{r0} a^{-4} + \rho_{d0} a^{-3}\ ,\quad p=\frac{\rho_{r0}}{3}a^{-4}\ ,
\ee
with constants $\rho_{r0}$ and $\rho_{d0}$. If the scale factor $a$
is given by a proper function $g(t)$ as
\be
\label{PQR10}
a=a_0\e^{g(t)}\ ,
\ee
with a constant $a_0$, Eq.(\ref{PQR7}) reduces
to the second rank differential equation (see also \cite{CNOT}):
\bea
\label{PQR11}
0&=&2 \frac{d^2 P(\phi)}{d\phi^2} - 2 g'(\phi) \frac{dP(\phi))}{d\phi} + 4g''(\phi) P(\phi) \nn
&& + \sum_i \left(1 + w_i\right) \rho_{i0} a_0^{-3(1+w_i)} \e^{-3(1+w_i)g(\phi)} \ .
\eea
In principle, by solving (\ref{PQR11}) we find the form of $P(\phi)$.
 Using
(\ref{PQR6}) (or equivalently (\ref{PQR7})), we also find the form of $Q(\phi)$ as
\bea
\label{PQR12}
Q(\phi)&=&-6 \left(g'(\phi)\right)^2 P(\phi)
 - 6g'(\phi) \frac{dP(\phi)}{d\phi} \nn
&& + \sum_i \rho_{i0} a_0^{-3(1+w_i)}  \e^{-3(1+w_i)g(\phi)} \ .
\eea
Hence, in principle, any cosmology
expressed as (\ref{PQR10}) can be realized by some specific
 $f(R)$-gravity.

\subsection{Exactly solvable example I: unification of matter dominated and
accelerated phases}

As an example, we consider the case
\be
\label{PQR13}
g'(\phi)=g_0 + \frac{g_1}{\phi}\ ,
\ee
 without matter $\rho=p=0$ for simplicity.  Eq.(\ref{PQR11}) reduces as
\be
\label{PQR14}
0=\frac{d^2 P}{d\phi^2} - \left(g_0 + \frac{g_1}{\phi}\right)\frac{dP}{d\phi}
   - \frac{2g_1}{\phi^2}P\ ,
\ee
whose solutions are given by the Kummer functions
(hypergeometric function of confluent type) as \cite{CNOT}
\be
\label{PQR15}
P=z^\alpha F_K(\alpha,\gamma; z)\ , \quad
z^{1-\gamma} F_K(\alpha - \gamma + 1, 2 - \gamma; z)\ .
\ee
Here
\bea
\label{PQR16} &&
z\equiv g_0\phi \ ,\quad \alpha\equiv \frac{1+g_1 \pm \sqrt{g_1^2 + 2g_1 + 9}}{4}\ ,\nn
&& \gamma\equiv 1\pm \frac{\sqrt{g_1^2 + 2g_1 + 9}}{2}\ ,
\eea
and the Kummer function is defined by
\be
\label{PQR17}
F_K(\alpha,\gamma;z)=\sum_{n=0}^\infty
\frac{\alpha(\alpha + 1)\cdots (\alpha + n -1)}{\gamma(\gamma + 1)\cdots
(\gamma + n - 1)} \frac{z^n}{n!}\ .
\ee
Eq.(\ref{PQR13})
tells that the Hubble rate $H$ is given by
\be
\label{PQR18}
H=g_0 + \frac{g_1}{t}\ .
\ee
When $t$ is small, as $H\sim g_1/t$,
the universe behaves as the one filled with a perfect fluid with the EOS
parameter
$w=-1 + 2/3g_1$. On the other hand when $t$ is large, $H$
approaches to constant $H\to g_0$ and the universe looks as
deSitter space. This shows the possibility of the transition from
matter dominated phase to the accelerating phase (compare with
\cite{CNOT}).
We should note that in this case, there is no matter and $f(R)$-terms
contribution plays the role of the matter
instead of the real matter. We will investigate later (in the next
subsection) the example that there is a real matter.
Similarly, one
can construct modified gravity action describing other epochs
bearing in mind that form of the modified gravity action is
different at different epochs (for instance, in inflationary epoch
it is different from the form at late-time universe).

We now investigate the asymptotic forms of $f(R)$ in (\ref{PQR4})
corresponding to (\ref{PQR13}). When $\phi$ and therefore $t$ are
small, we find
\be
\label{PQR19}
P\sim P_0\phi^\alpha\ ,\quad
Q\sim -6P_0g_1\left(g_1 + \alpha\right)\phi^{\alpha - 2}\ .
\ee
Here $P_0$ is a constant. Using (\ref{PQR2}), it follows
\be
\label{PQR20}
\phi^2 \sim \frac{6g_1\left(g_1 + \alpha\right)\left(\alpha - 2\right)}{\alpha R}\ ,
\ee
which gives
\be
\label{PQR21}
f(R) \sim - \frac{2P_0}{\alpha - 2}\left\{
\frac{6g_1\left(g_1 + \alpha\right)\left(\alpha
 - 2\right)}{\alpha}\right\}^{\alpha/2} R^{1 - \frac{\alpha}{2}}\ .
\ee
On the other hand, when $\phi$ and therefore $t$ are positive
and large, one gets
\bea
\label{PQR22}
P&\sim & \tilde P_0 \phi^{2\alpha - \gamma}\e^{g_0\phi}\left(1
+ \frac{(1-\alpha)(\gamma - \alpha)}{g_0\phi}
\right) \nn
Q&\sim & - 12 g_0^2 \tilde P_0 \phi^{2\alpha - \gamma}\e^{g_0\phi} \nn
&& \times \left(1 + \frac{-9 + 12 \alpha
 - 5\gamma - 2\alpha \gamma + 2\alpha^2}{2g_0\phi}
\right) \nn
\phi&\sim& \frac{-\frac{9}{2} + 9\alpha
 - \frac{7}{2}\gamma}{g_0\left(\frac{R}{12g_0^2} - 1\right)}\ .
\eea
Here $\tilde P_0$ is a constant. Then we find
\bea
\label{PQR23}
f(R)&\sim& 12 g_0^2 \tilde P_0 \left\{\frac{1}{g_0}
\left( -\frac{9}{2} + 9\alpha - \frac{7}{2}\gamma
\right)\right\}^{2\alpha - \gamma} \nn
&& \times
\left(\frac{R}{12g_0^2} - 1\right)^{-2\alpha + \gamma + 1} \nn
&& \times \exp\left( \frac{ -\frac{9}{2} + 9\alpha
 - \frac{7}{2}\gamma }{\frac{R}{12g_0^2} - 1}\right)\ .
\eea
This shows the principal possibility of unification of matter-dominated
phase (even without matter!), transition to acceleration and late-time
speed up of the universe
for specific, implicitly given model of $f(R)$ gravity.

\subsection{Exactly solvable example II: model reproducing $\Lambda$CDM-type cosmology}

Let us investigate if $\Lambda$CDM-type cosmology could be reproduced by
$f(R)$-gravity in the
present formulation.

In the Einstein gravity, when there is a matter with the EOS parameter $w$
and cosmological constant,
the FRW equation has the following form:
\be
\label{LCDM1}
\frac{3}{\kappa^2}H^2 = \rho_0 a^{-3(1+w)} + \frac{3}{\kappa^2 l^2}\ .
\ee
Here $l$ is the length parameter coming from the cosmological constant.
The solution of (\ref{LCDM1}) is given by
\bea
\label{LCDM2}
a&=&a_0\e^{g(t)}\ ,\nn
g(t)&=&\frac{2}{3(1+w)}\ln \left(\alpha \sinh \left(\frac{3(1+w)}{2l}\left(t - t_0 \right)\right)\right)\ .
\eea
Here $t_0$ is a constant of the integration and
\be
\label{LCDM3}
\alpha^2\equiv \frac{1}{3}\kappa^2 l^2 \rho_0 a_0^{-3(1+w)}\ .
\ee
It is possible to reconstruct $f(R)$-gravity reproducing (\ref{LCDM2}).
When  the matter contribution is neglected , Eq.(\ref{PQR11})  has the
following form:
\bea
\label{LCDM4}
0 &=& 2\frac{d^2 P(\phi)}{d\phi^2} - \frac{2}{l}\coth \left(\frac{3(1+w)}{2l}\left(t - t_0 \right)\right) \frac{dP(\phi)}{d\phi} \nn
&& - \frac{6(1+w)}{l^2} \sinh^{-2} \left(\frac{3(1+w)}{2l}\left(t - t_0 \right)\right) P(\phi)\ .
\eea
By changing the variable from $\phi$ to $z$ as follows,
\be
\label{LCDM5}
z\equiv - \sinh^{-2} \left(\frac{3(1+w)}{2l}\left(t - t_0 \right)\right) \ ,
\ee
Eq.(\ref{LCDM4}) can be rewritten in the form of Gauss's hypergeometric differential equation:
\bea
\label{LCDM6}
&& 0=z(1-z)\frac{d^2 P}{dz^2} + \left[\tilde\gamma - \left(\tilde\alpha + \tilde \beta + 1\right)z\right] \frac{dP}{dz}
 - \tilde\alpha \tilde\beta P\ , \nn
&& \tilde\gamma \equiv 4 + \frac{1}{3(1+w)} \ ,\quad
\tilde\alpha + \tilde\beta + 1 \equiv 6 + \frac{1}{3(1+w)}\ ,\nn
&& \tilde\alpha \tilde\beta \equiv - \frac{1}{3(1+w)}\ ,
\eea
whose solution is given by Gauss's hypergeometric function:
\bea
\label{LCDM7}
P&=& P_0 F(\tilde\alpha,\tilde\beta,\tilde\gamma;z) \nn
&\equiv& P_0 \frac{\Gamma(\tilde\gamma)}{\Gamma(\tilde\alpha) \Gamma(\tilde\beta)}
\sum_{n=0}^\infty \frac{\Gamma(\tilde\alpha + n) \Gamma(\beta + n)}{\Gamma(\tilde\gamma + n)}
\frac{z^n}{n!}\ .
\eea
Here $\Gamma$ is the $\Gamma$-function. There is one more linearly independent solution like
$(1-z)^{\tilde\gamma - \tilde\alpha - \tilde\beta}F(\tilde\gamma - \tilde\alpha, \tilde\gamma - \tilde\beta, \tilde\gamma;z)$
but we drop it, for simplicity.
Using (\ref{PQR12}), one finds the form of $Q(\phi)$:
\bea
\label{LCDM8}
Q&=& - \frac{6(1-z)P_0}{l^2}F( \tilde\alpha,\tilde\beta,\tilde\gamma;z) \nn
&& - \frac{3(1+w) z(1-z)P_0}{l^2(13 + 12w)}
F(\tilde\alpha+1,\tilde\beta+1,\tilde\gamma+1;z)\ .
\eea
>From (\ref{LCDM5}), it follows $z\to 0$ when $t=\phi\to + \infty$. Then
in the limit, one arrives at
\be
\label{LCDM9}
P(\phi)R + Q(\phi) \to P_0 R - \frac{6P_0}{l^2}\ .
\ee
Identifying
\be
\label{LCDM10}
P_0=\frac{1}{2\kappa^2}\ ,\quad \Lambda = \frac{6}{l^2}\ ,
\ee
the Einstein theory with cosmological constant $\Lambda$ can be
reproduced.
The action is not singular even in the limit of $t\to \infty$.
Note that slightly different approach to construct $\Lambda$CDM cosmology
 from $f(R)$ gravity is developed in ref.\cite{dobado}.

Therefore even without cosmological constant nor cold dark matter,
the cosmology of $\Lambda$CDM model could be reproduced
by $f(R)$-gravity. It was shown in ref.\cite{abdalla} that some versions
of modified gravity
contain Big Rip singularities \cite{mcinnes} (for their classification,
see
\cite{tsujikawa}). Hence, the above model without future singularity and
with typical $\Lambda$CDM behaviour looks quite realistic.

\subsection{Models of $f(R)$ gravity with transition of matter dominated
phase  to the acceleration phase}

Let us consider more realistic examples where the total action contains
also usual matter.
The starting form of  $g(\phi)$ is
\be
\label{PQR24}
g(\phi)=h(\phi) \ln \left(\frac{\phi}{\phi_0}\right)\ ,
\ee
with a constant $\phi_0$. It is assumed  that $h(\phi)$ is a slowly
changing function of $\phi$.
We use adiabatic approximation and neglect the derivatives
of $h(\phi)$ $\left(h'(\phi)\sim h''(\phi) \sim 0\right)$.
Eq.(\ref{PQR11}) has the following form:
\bea
\label{PQR25}
0&=& \frac{d^2 P(\phi)}{d\phi^2} - \frac{h(\phi)}{\phi} \frac{dP(\phi))}{dt}
 - \frac{2h(\phi)}{\phi^2}  P(\phi) \nn
&& + \sum_i \rho_{i0} a_0^{-3(1+w_i)} \e^{-3(1+w_i)g(\phi)} \ .
\eea
The solution for $P(\phi)$ is found to be
\bea
\label{PQR26}
P(\phi) &=& p_+ \phi^{n_+(\phi)} + p_- \phi^{n_-(\phi)} \nn
&& + \sum_i p_i(\phi) \phi^{-3(1+w_i)h(\phi) + 2} \ .
\eea
Here $p_\pm$ are arbitrary constants and
\bea
\label{PQR27}
n_\pm (\phi) &\equiv& \frac{h(\phi) - 1 \pm \sqrt{h(\phi)^2 + 6h(\phi) + 1}}{2}\ ,\nn
p_i (\phi) &\equiv& - \left\{(1+w)\rho_{i0} a_0^{-3(1+w_i)} \phi_0^{3(1+w)h(\phi)}\right\} \nn
&& \times \left\{6(1+w)(4+3w) h(\phi)^2 \right. \nn
&& \left. \qquad - 2 \left(13 + 9w\right)h(\phi) + 4\right\}^{-1}\ .
\eea
Especially for the radiation and dust, one has
\bea
\label{PP2}
p_r (\phi) & \equiv & - \frac{4\rho_{r0}\phi_0^{4h(\phi)} }{3a_0^4
\left( 40 h(\phi)^2 - 32 h(\phi) + 4\right)}\ ,\nn
p_d (\phi) & \equiv & - \frac{\rho_{d0}\phi_0^{3h(\phi)} }{a_0^3
\left( 24 h(\phi)^2 - 26 h(\phi) + 4\right)}\ .
\eea
We also find the form of $Q(\phi)$ as
\bea
\label{PQR28}
Q(\phi) &=& - 6h(\phi)p_+ \left(h(\phi) + n_+(\phi) \right) \phi^{n_+ (\phi) - 2} \nn
&& - 6h(\phi)p_- \left(h(\phi) + n_-(\phi) \right) \phi^{n_- (\phi) - 2} \nn
&& + \sum_i\left\{ - 6h(\phi) \left( -(2+3w)h(\phi) + 2\right)p_i (\phi) \right. \nn
&& \left.  + p_{i0} a_0^{-3(1+w)}\phi_0^{3(1+w)h(\phi)}\right\} \phi^{- 3(1+w)h(\phi)} \ .
\eea
Eq.(\ref{PQR24}) tells that
\be
\label{PQR29}
H\sim\frac{h(t)}{t}\ .
\ee
and
\be
\label{PQE56}
R\sim \frac{6\left( -h(t) + 2h(t)^2\right)}{t^2}\ .
\ee
Let assume $\lim_{\phi\to 0} h(\phi) = h_i$ and $\lim_{\phi\to \infty} h(\phi) = h_f$.
Then if $0<h_i<1$, the early universe is in decceleration phase and
if $h_f>1$, the late universe is in acceleration phase.
We may consider the case $h(\phi)\sim h_m$ is almost constant when $\phi\sim t_m$
$\left(0\ll t_m \ll +\infty\right)$.
If $h_1$, $h_f>1$ and $0<h_m<1$,  the early universe is also accelerating, which
could be inflation. After that the universe becomes deccelerating,
which corresponds
to matter-dominated phase with $h(\phi)\sim 2/3$ there.
Furthermore,
after that, the universe could be in the acceleration phase.

The simplest example is
\be
\label{PQ1}
h(\phi) = \frac{h_i + h_f q \phi^2}{1 + q \phi^2}\ ,
\ee
with constants $h_i$, $h_f$, and $q$. When $\phi\to 0$, $h(\phi)\to h_i$ and
when $\phi\to \infty$, $h(\phi)\to h_f$. If $q$ is small enough, $h(\phi)$ can be a slowly
varying function of $\phi$.
By using the expression of (\ref{PQE56}), we find
\bea
\label{PQ2}
&& \phi^2=\Phi_0(R)\ ,\quad \Phi_\pm (R)\ ,\nn
&& \Phi_0 \equiv \alpha_+^{1/3} + \alpha_-^{1/3}\ ,\quad
\Phi_\pm \equiv \alpha_\pm^{1/3}\e^{2\pi i/3} + \alpha_\mp^{1/3}\e^{-2\pi i/3}\ ,\nn
&& \alpha_\pm \equiv \frac{-\beta_0 \pm \sqrt{\beta_0^2 - \frac{4\beta_1^3}{27}}}{2}\ ,\nn
&& \beta_0 \equiv \frac{2\left(2R + 6h_f q - 12 h_f^2 q \right)^3}{27 q^3 R^3} \nn
&& \qquad - \frac{\left(2R + 6h_f q - 12 h_f^2 q \right)\left(R + 6h_i q + 6 h_f q - 4h_i h_f q\right)}{3qR} \nn
&& \qquad + 6h_i - 12 h_i^2\ ,\nn
&& \beta_1 \equiv - \frac{\left(2R + 6h_f q - 12 h_f^2 q \right)^2}{3 q^2 R^2} \ ,\nn
&& \qquad - \frac{R + 6h_i q + 6 h_f q - 4h_i h_f q}{q^2 R}\ .
\eea
There are three branches $\Phi_0$ and $\Phi_\pm$ in (\ref{PQ2}).
 Eqs.(\ref{PQE56}) and (\ref{PQ1}) show that when the curvature
is small ($\phi=t$ is large), we find $R\sim 6\left( - h_f + 2 h_f \right)/\phi^2$ and when the curvature is large ($\phi=t$
is small), $R\sim 6\left( - h_i + 2 h_i \right)/\phi^2$.
 This asymptotic behaviour indicates that we should choose $\Phi_0$ in
(\ref{PQ2}). Then explicit form of $f(R)$ could be given by using
the expressions of $P(\phi)$  (\ref{PQR26}) and
$Q(\phi)$  (\ref{PQR28}) as
\be
\label{PQ3}
f(R)=P\left(\sqrt{\Phi_0(R)}\right) R + Q\left(\sqrt{\Phi_0(R)}\right)\ .
\ee

One may check the asymptotic behavior of $f(R)$ in (\ref{PQ3}).
For simplicity, it is considered the case that the matter is only dust
($w=0$) and  that $p_-=0$.
Then we find
\be
\label{nm1}
P(\phi) = p_+ \phi^{n_+(\phi)} + p_d(\phi) \phi^{-3h(\phi) + 2} \ .
\ee
One may always get
\be
\label{nm2}
n_+ - \left(-3h + 2\right) >0\ .
\ee
in (\ref{nm1}). Here $n_+=\left(h(\phi) - 1 \pm \sqrt{h(\phi)^2 + 6h(\phi) + 1}\right)/2$ is defined in (\ref{PQR27}).
Then when $\phi$ is large, the first term in (\ref{nm1}) dominates and
when $\phi$ is small, the last term dominates.
When $\phi$ is large, curvature is small and  $\phi^2\sim 6\left( - h_f + 2 h_f \right)/R$
and $h(\phi)\to h(\infty)=h_f$.
Hence, Eq.(\ref{nm1}) shows that
\be
\label{PQasym1}
P(\phi) \sim p_+ \left(\frac{6\left( - h_f + 2 h_f \right)}{R}\right)^{\left(h_f - 1
+ \sqrt{h_f^2 + 6h_f + 1}\right)/4} \ ,
\ee
and therefore
\be
\label{PQasym2}
f(R)\sim R^{-\left(h(\phi) - 5 + \sqrt{h_f^2 + 6h_f + 1}\right)/4 } \ .
\ee
Especially when $h\gg 1$, we find
\be
\label{PQasym2b}
f(R)\sim R^{-h_f/2}\ .
\ee
Therefore there appears the negative power of $R$. As $H\sim h_f/t$, if $h_f>1$, the universe is in acceleration
phase.

On the other hand, when curvature is large, we find
$\phi^2\sim 6\left( - h_i + 2 h_i \right)/R$ and $h(\phi)\to h(0)=h_i$.
Then (\ref{PQR26}) shows
\be
\label{PQasym3}
P(\phi) \sim p_d(0) \phi^{-3h_i + 2} \ .
\ee
If the universe era corresponds to matter dominated phase ($h_i=2/3$),
$P(\phi)$ becomes a constant and therefore
\be
\label{PQasym4}
f(R)\sim R\ ,
\ee
which reproduces the Einstein gravity.

Thus, in the above model, matter dominated phase evolves into acceleration
phase and $f(R)$ behaves as
$f(R)\sim R$ initially while $f(R)\sim R^{-\left(h(\phi) - 5 + \sqrt{h_f^2
+ 6h_f + 1}\right)/4 }$ at late time.

Three years WMAP data are recently analyzed in Ref.\cite{Spergel}, which
shows that the combined analysis of WMAP with supernova Legacy
survey (SNLS) constrains the dark energy equation of state $w_{DE}$ pushing it
towards the cosmological constant. The marginalized best fit values of the
equation of state parameter at 68$\%$ confidance level
are given by $-1.14\leq w_{\rm DE} \leq -0.93$. In case of a prior that universe is
flat, the combined data gives  $-1.06 \leq w_{\rm DE} \leq -0.90 $.

In our model, we can identify
\be
\label{w1}
w_{\rm DE} = -1 + \frac{2}{3 h_f}\ ,
\ee
or
\be
\label{w2}
h_f= \frac{2}{3\left(1 + w_{\rm DE}\right)}\ ,
\ee
which tells $h_f$ should be large if $h_f$ is positive. For example, if $w_{\rm DE}=-0.93$, $h_f\sim 9.51\cdots$ and
if $w_{\rm DE}=-0.90$, $h_f\sim 6.67\cdots$.
Thus, we presented the example of $f(R)$ gravity which describes the
matter dominated stage, transition from decceleration to acceleration and
acceleration epoch which is consistent with three years WMAP.

\subsection{(In)stability of the cosmological solutions}

Let us investigate the stability of the obtained solutions.
We assume
\be
\label{PQR30}
a=a_0\e^{g(\phi)}\ ,
\ee
which corresponds to (\ref{PQR10}) and $P(\phi)$ should be given by a solution of (\ref{PQR11})
(and $Q(\phi)$ should be given by (\ref{PQR12})).
Under the above assumptions, we consider the perturbations in (\ref{PQR6})
and (\ref{PQR7}).
By deleting $Q(\phi)$ in (\ref{PQR6}) and (\ref{PQR7}), one obtains
\bea
\label{PQR31}
0&=&2 \frac{d^2 P(\phi)}{dt^2} - 2 g'(\phi) \frac{dP(\phi))}{dt} + 4g''(\phi) P(\phi) \nn
&& + \sum_i ( 1+w_i )\rho_{i0} a_0^{-3(1+w_i)} \e^{-3(1+w_i)g(\phi)} \nn
&=&2 \frac{d^2 P(\phi)}{d\phi^2} \left(\frac{d\phi}{dt}\right)^2
+ 2 \frac{d P(\phi)}{d\phi} \frac{d^2\phi}{dt^2} \nn
&& - 2 g'(\phi) \frac{dP(\phi))}{d\phi}\left(\frac{d\phi}{dt}\right)^2 \nn
&& + 4\left\{g''(\phi) \left(\frac{d\phi}{dt}\right)^2 + g'(\phi) \frac{d^2\phi}{dt^2} \right\}P(\phi) \nn
&& + \sum_i ( 1+w_i )\rho_{i0} a_0^{-3(1+w_i)} \e^{-3(1+w_i)g(\phi)} \ .
\eea
Then combining (\ref{PQR11}) with (\ref{PQR30}) it follows
\bea
\label{PQR32}
0&=&2 \left\{ \frac{d^2 P(\phi)}{d\phi^2}  - g'(\phi) \frac{dP(\phi))}{d\phi}
+ 2 g''(\phi) P(\phi) \right\} \nn
&&\times \left\{ \left(\frac{d\phi}{dt}\right)^2 - 1\right\} \nn
&& + 2 \left\{ \frac{d P(\phi)}{d\phi} + 2 g'(\phi) P(\phi)\right\}\frac{d^2\phi}{dt^2} \ .
\eea
By defining $\delta$ as
\be
\label{PQR33}
\delta \equiv \frac{d\phi}{dt} - 1\ ,
\ee
we consider the perturbation from the solution (\ref{PQR8}).
 Using (\ref{PQR32}), one gets
\bea
\label{PQR34}
\frac{d\delta}{dt}&=&- \omega(t)\delta\ ,\\
\omega(t)&\equiv& \left. 2\frac{\frac{d^2 P(\phi)}{d\phi^2}  - g'(\phi) \frac{dP(\phi))}{d\phi}
+ 2 g''(\phi) P(\phi)}{\frac{d P(\phi)}{d\phi} + 2 g'(\phi) P(\phi)}\right|_{\phi=t}\ .\nonumber
\eea
Then when $\omega>0$ ($\omega<0$), the perturbation becomes small (large) and the system is
stable (unstable).

As an example,  the case  (\ref{PQR24}) may be considered. It gives
(\ref{PQR26}) with (\ref{PQR27}).
Especially when $p_\pm=p_{i0}=0$ except $p_{di}$, we find
\be
\label{PQR35}
\omega=\frac{2\left(12h^2 - 13 h + 2\right)}{(2-h)t}\ .
\ee
Here  the derivatives of $h(\phi)$ like $h'(\phi)$ are neglected again.
Then $\omega$ goes to infinity when $h=2$ and $h\to \pm\infty$ and
$\omega$ vanishies when
\be
\label{PQR36}
h=h_\pm = \frac{13\pm \sqrt{63}}{24}=0.87238\cdots,\ 0.2109478\cdots\ .
\ee
Therefore $\omega>0$ and the system is stable when $h<h_-$ or $h_+<h<2$.
 Hence, for  the case that
the universe starts from decceleration phase with $h=h_0<1$, if $h_0>h_+$, there is a stable solution where universe
develops to the acceleration phase $h\to 2>1$.
Even if we started with $h(\phi)=2/3$, which corresponds to matter
dominated phase $a\sim t^{2/3}$, the solution is
unstable since $h_-<2/3<h_+$, the perturbed solution could develop into the stable solution with $h>h_+$ and
therefore there could be a transition into the acceleration phase. If $h$
goes to $2$ from the region with $h<2$,
since $\omega\to +\infty$, the solution becomes extremely stable. Hence,
$h$ may pass through the point $h=2$ and
$h$ could become larger than $2$, where the effective EOS parameter $w=1 -
2/3h=7/9$.

Note that when $p_\pm=p_{i0}=0$ $\left(i\neq d:\mbox{dust}\right)$, one
 gets
\be
\label{PQR37}
R\sim \phi^{-2}\ ,\quad f(R) \sim R^{3h/2}\ .
\ee
Therefore in the matter dominated phase $h\sim 2/3$, the action behaves as
the Hilbert-Einstein action.

In more general case that there is only one kind of matter with $w$
and $p_\pm=0$, we find
\bea
\label{PP3}
R&=& \frac{3h(\phi)\left\{12 \left( 1 + w \right) h(\phi)^2 - 2 \left(7 + 9w \right) h(\phi) + 4 \right\}}
{\left\{3\left(1 + w \right) h(\phi) + 2\right\} \phi^2}\ ,\nn
f(R)& \sim & R^{\frac{3}{2}\left(1+w\right)h(\phi(R))}\ .
\eea
Note that $\omega$ in (\ref{PQR34}) is given by
\be
\label{PP4}
\omega= \frac{3(1+w)(4+3w) h(\phi)^2 - (13 + 9w)h(\phi) + 2}{\left\{ - (1 + 3w) h(\phi) + 2 \right\} t}\ .
\ee

Next we consider the case that $p_{d0}\neq 0$ and $p_{r0}\neq 0$ but $p_\pm=0$ and $p_{i0}=0$
except $i\neq d, r$.  Then it follows
\bea
\label{PQR38}
\omega&=&- \frac{1}{\phi}\left\{ 4p_r \left(10 h^2 - 8h + 1\right) +2p_d \left( 12 h^2 - 13 h + 2 \right)\phi^h \right\} \nn
&& \times \left\{2p_r\left(h-1\right) + p_d\left(h - 2\right)\phi^h \right\}^{-1} \ ,
\eea
or by using (\ref{PQR27}),
\bea
\label{PQR39}
\omega&=& - \frac{2}{\phi}\left(\frac{4\rho_{r0} \phi_0^{4h}}{3a_0^4} + \frac{\rho_{d0} \phi_0^{3h}\phi^h}{a_0^3}\right)
\left(10h^2 - 8h + 1\right) \nn
&& \times \left(12h^2 - 13 h + 2\right) \nn
&& \times \left\{ \frac{4\rho_{r0} \phi_0^{4h}}{3a_0^4} \left(12h^2 - 13 h + 2\right) \left( h - 1\right) \right. \nn
&& \left. + \frac{\rho_{d0} \phi_0^{3h}}{a_0^3} \phi^h \left(10h^2 - 8h + 1\right) \left( h -2 \right)\right\}^{-1}\ .
\eea
As clear from (\ref{PQR9}), (\ref{PQR24}), and (\ref{PQR30}), $\rho_{r0} 3a_0^{-4}$ and
$\rho_{d0} a_0^{-3}$ correspond to the energy density for radiation and dust (usual matter plus
cold dark matter), respectively, when $t=\phi=\phi_0$. Let us choose
$t=\phi_0$ corresponding to the present universe.
It  may be assumed
\be
\label{PQR40}
\epsilon \equiv \left(\frac{4\rho_{r0} \phi_0^{4h}\phi^h }{3a_0^4}\right)
\left(\frac{\rho_{d0} \phi_0^{3h}}{a_0^3}\right)^{-1}\ll 1\ .
\ee
In the expression (\ref{PQR39}), $\omega$ vanishes when $h=h_\pm$  (\ref{PQR36}) and $h=\tilde h_\pm$, defined by
\be
\label{PQR41}
\tilde h_\pm \equiv \frac{4\pm \sqrt{6}}{10}=0.6449\cdots, 0.15505\cdots\ .
\ee
Under the assumption (\ref{PQR40}), $\omega$ diverges at
\bea
\label{PQR42}
&& h=2 - \frac{24}{25}\epsilon\ ,\quad
h=\tilde h_\pm + \epsilon \delta_\pm\ ,\nn
&& \delta_\pm \equiv \mp \frac{\left(28\pm 17\sqrt{6}\right)\left( -6 \pm \sqrt{6}\right)
\left(16\pm \sqrt{6}\right)}{25000}\ .
\eea
Here $\delta_\pm>0$. Near the singularities, if $\omega$ is negative,
there could be very large instability, which
should be avoided. In case $p_\pm \neq 0$, the singularity could be
avoided. The points where $\omega$ vanishes could
remain even if $p_\pm \neq 0$ but the instability becomes finite and
there can be a solution which describes the transition from
the matter dominated phase to the acceleration phase.

We now consider the case that the contribution from the matter could be neglected in (\ref{PQR26}) and therefore
we could assume that $p_{i0}=0$. Furtheremore when one of $p_\pm$ vanishes,
 one finds
\be
\label{PQR43}
R \sim \phi^{-2}\ ,\quad f(R)\sim R^{N_\pm}\ ,\quad N_\pm \equiv 1 - \frac{n_\pm}{2}\ .
\ee
The behavior of $N_\pm$ is as following
\bea
\label{PQR44}
\mbox{when} & h \to 0\ ,\quad & N_+ \to 1\ \mbox{and}\ N_- \to \frac{3}{3} \nn
& h \to + \infty\ , & N_+ \to - \frac{h}{2}\ \mbox{and}\ N_-\to 2 \nn
& h=1\ , & N_\pm \to 1 \mp \frac{1}{\sqrt{2}} \nn
& h=\frac{2}{3}\ , & N_+ \to \frac{1}{3}\ \mbox{and}\ N_-\to \frac{3}{2}\ .
\eea
Thus, in case $p_+=0$ but $p_-\neq 0$, when $h\to + \infty$, the higher
derivative inflationary  model $f(R)\sim R^2$ appears.
We should also note $\omega$ in (\ref{PQR34}) has the following form:
\be
\label{PQR45}
\omega = \omega_\pm \equiv - \frac{2\left(h(\phi) - 1 \right) \pm 4 \sqrt{h(\phi)^2 + 6h (\phi) + 1}}
{5h(\phi) - 1 \pm \sqrt{h(\phi)^2 + 6h (\phi) + 1}}\ .
\ee
Now $\hat h_\pm$ may be defined as
\be
\label{PQE46}
\hat h_\pm =\frac{13 \pm 4 \sqrt{10}}{3}=8.5497\cdots, 0.11690\cdots\ .
\ee
Then  $\omega_+<0$ when $h>\hat h_+$ or $h<0$, and $\omega_+>0$ when $0<h<\hat h_+$.
We also find that when $\omega_->0$, $h>2/3$ or $h<h_-$, and $\omega_-<0$ when $\hat h_- < h < 2/3$.
Hence, the model where $p_+=0$ but $p_-\neq 0$ is stable when
$h>\frac{2}{3}$. Thus,
one can make a stable model where matter dominated phase $h\sim 3/2$
evolves to the acceleration
phase $h>1$.

We should note that $p_-=0$ and $p_{i0}=0$ but $p_+\neq 0$, if we put
\be
\label{PQE47}
h=\frac{10}{3}\ ,
\ee
we obtain $N_+=-1$ or $f(R)\sim 1/R$. Let us put
\be
\label{PQE47b}
h(\phi)=\frac{10}{3} + \delta h\ ,\quad \left|\delta h\right|\ll 1\ .
\ee
It follows
\be
\label{PQE48}
N_+ = -1 -\frac{18}{17}\delta h\ .
\ee
When $h\sim 10/3$, from (\ref{PQR2}), the curvature is given as
\be
\label{PQE49}
R\sim \frac{6h(h+n_+)(n_+ - 2)}{n_+ \phi^2}\sim \frac{220}{3\phi^2}\ .
\ee
Hence, with the choice
\bea
\label{PQE50}
\delta h &\sim& - \frac{17}{18\ln \frac{220}{3\mu^2\phi^2}}\left(\frac{220^2}{18\kappa^2\mu^6\phi^4}
+ \frac{220^3 \beta}{27 \mu^6 \phi^6}\right) \nn
&\sim& - \frac{17}{18\ln \frac{R}{\mu^2}}\left(\frac{R^2}{2\kappa^2\mu^6}
+ \frac{\beta R^3}{\mu^6}\right)\ ,
\eea
with a constant $\mu$, which has a dimension of mass, one arrives at
\be
\label{PQE51}
f(R)\sim \frac{\mu^6}{R} + \frac{R}{2\kappa^2} + \beta R^2\ ,
\ee
which reproduces the action proposed in \cite{NO}.
(Note that such class of actions does not describe sequence of matter
dominated/acceleration phase \cite{david}).
Thus, using stability analysis we demonstrated that indeed the matter
dominated phase may transit to the acceleration phase for some implicit
model
of $f(R)$ gravity found in the previous subsection.
Moreover, it is shown that the model of ref.\cite{NO} with positive and
negative powers of the curvature is just asymptotic form of such
consistent
$f(R)$ theory (at some specific values of parameters) at acceleration
epoch. The complete, implicit version of
$f(R)$ theory found in previous subsection  decribes the sequence of
matter dominated phase, transition from deceleration to acceleration and
then acceleration epoch of the universe.

\subsection{ Corrections to Newton law}

In the present subsection we will discuss the contributions to Newton law
in the modified gravity under consideration. Note that there is big number
of papers devoted to the study of newtonian regime in modified $f(R)$
gravity \cite{newton,NO}. However, these papers are mainly devoted to the
study of newtonian regime for   $1/R$ models.
We now check when the corrections to the Newton law are not essential in
 $f(R)$-gravity under consideration.
For this purpose, we put a point source at $\bmr=0$:
\be
\label{New1}
\rho_m = \frac{m}{a(t)^3}\delta (\bmr)\ .
\ee
Transforming
\be
\label{New2}
g_{\mu\nu}\to g_{\mu\nu} + \delta g_{\mu\nu}\ ,\quad
\phi\to \phi + \delta \phi\ ,
\ee
one finds the $(t,t)$-component of (\ref{PQR5}) has the following form:
\bea
\label{New3}
0 &=& G_0 + G_1 \ ,\nn
G_0 &\equiv & - \frac{1}{2}\delta g_{tt} \left\{ P(\phi) \left(6\dot H + 12 H^2\right) + Q(\phi) \right\} \nn
&& - \frac{1}{2}P(\phi) \delta R - P(\phi) \delta R_{tt} \ ,\nn
G_1 &=& \frac{1}{2}\left\{P'(\phi) \left( 6\dot H + 12 H^2 \right) + Q'(\phi)\right\} \delta \phi \nn
&& + 3\left(\dot H + H^2 \right) P'(\phi) \delta \phi \nn
&& + \frac{1}{2} P'(\phi) \nabla_t \delta g_{tt} + 3H P'(\phi) \delta g_{tt} + P''(\phi) \delta g_{tt} \nn
&& - H P'(\phi) g^{ij} \delta g_{ij} + \frac{1}{2} P'(\phi) \nabla_t \left(g^{\mu\nu} \delta g_{\mu\nu}\right) \nn
&& + \frac{m}{2a(t)^3}\delta (\bmr)\ .
\eea
Here  the gauge condition is chosen
\be
\label{New4}
\nabla^\mu \delta g_{\mu\nu}=0\ .
\ee
On the other hand, Eq.(\ref{PQR2}) gives
\be
\label{New5}
0=\left( P''(\phi) + Q''(\phi) R \right) \delta \phi + Q'(\phi) \delta R\ .
\ee
We now consider the region for $r=|\bmr|$ as
\be
\label{New6}
\frac{1}{m}\ll r \ll \frac{1}{H}\ ,
\ee
or
\be
\label{New7}
m\gg \frac{\partial}{\partial r}\gg H\ .
\ee
We should also note that
\be
\label{New8}
\frac{\partial}{\partial t} \sim \frac{1}{\phi} \sim H \sim \sqrt{R}\ .
\ee
Then if
\be
\label{New9}
P'(\phi)\ll H P(\phi)\ ,\quad P''(\phi)\ll H^2 P(\phi)\ ,
\ee
$G_1$  (\ref{New3}) could be neglected if compared with $G_0$.
Then since $\phi$ can be regarded as a constant and therefore $\delta\phi=0$ as long as (\ref{New6})
is satisfied, Eq.(\ref{New3}) reduces to that in the Einstein gravity by identifying $P(\phi)$ with
$1/\kappa^2$ and $Q(\phi)=\Lambda^2/\kappa^2$ since $P(\phi)$ and $Q(\phi)$
 are slowly varying (almost constant) functions.
From
(\ref{New5}),
 $\delta \phi=0$ implies $\delta R=0$. Then Eq.(\ref{New3}) gives $(t,t)$-component of
the usual perturbation in the Einstein equation:
\bea
\label{New10}
0&=&G_0 \nn
&=& - \frac{1}{2\kappa^2}\delta g_{tt} \left\{ R_0 + \Lambda^2 \right\}
 - \frac{1}{\kappa^2}\left(\frac{1}{2} \delta R - \delta R_{tt}\right) \nn
&& + \frac{m}{2a_0^3}\delta (\bmr) \nn
&& R_0\equiv 6\dot H + 12 H^2 \ .
\eea
Here $a_0$ is the scale factor in the present universe $a_0=a(t)$, which may be chosen to be
unity $a_0=1$.
For the almost flat universe as our current one, we can neglect the
first term in
(\ref{New10}): $R_0 + \Lambda^2 \sim 0$ and we obtain
\be
\label{New11}
0 = - \frac{1}{\kappa^2}\left(\frac{1}{2} \delta R - \delta R_{tt}\right) + m\delta (\bmr) \ ,
\ee
which further reduces, since the universe is almost flat and the source and the universe are static
in a region where we are investigating the Newton law, to
\be
\label{New12}
0 = \frac{1}{2\kappa^2}\left( \triangle \delta g_{tt} - \triangle \left(g^{\mu\nu}\delta g_{\mu\nu}\right)\right)
+ m\delta (\bmr) \ .
\ee
Here $\triangle$ is usual Laplacian.
Therefore if the conditions in (\ref{New8}) are satisfied, the correction to the Newton law could be
neglected.
In case of (\ref{PQR29}), if we choose $p_+$ and $p_-$ to satisfy (\ref{New9}) in the present universe,
the correction to the Newton law could be small. (It is interesting to
note that standard Newton law is valid also in an arbitrary $F(G)$ gravity
\cite{cognola} where $G$ is Gauss-Bonnett invariant).
One may consider even simpler situation: admitting that Newton law is
satisfied only in current universe.
Then, the form of Newton law should be fixed only at present universe
(in other words, the initial value of $f(R)$ should be restricted).
We should also note that if $P(\phi)$ changes its sign, as we identified $P(\phi)$ with $1/\kappa^2$,
there could appear the anti-gravitation effect. For instance, the form of
modified gravity may be changed in the future, at the end of acceleration
epoch, driving the Newton law to its opposite sign form.

\section{Modified gravity and compensating dark energy}

In the present section we will present another approach to modified
gravity.
Specifically, we discuss the modified gravity which successfully
describes the acceleration epoch but may be not viable in matter dominated
stage. In this case, it is demonstrated that one can introduce the
compensating dark energy (some ideal fluid) which helps to
realize matter dominated and decceleration-acceleration transition phases.
The role of such compensating dark energy is negligible in the
acceleration epoch.

We now start with general $f(R)$-gravity action:
\be
\label{PQE52}
S=\int d^4 x \left\{f(R) + {\cal L}_{\rm matter}\right\}\ .
\ee
In the FRW metric with flat spatial dimensions one gets
\bea
\label{PQE53}
\rho &=& f(R) - 6\left(\dot H + H^2 - H \frac{d}{dt}\right)f'(R)\ ,\nn
p&=& - f(R) - 2 \left( - \dot H - 3 H^2 + \frac{d^2}{dt^2} + 2 H \frac{d}{dt}\right) f'(R)\ ,\nn
R&=& 6\dot H + 12 H^2\ .
\eea
If the Hubble rate is given (say, by observational data) as a function of
$t$: $H=H(t)$,
by substituting such expression into (\ref{PQE53}), we find the
$t$-dependence of $\rho$ and $p$ as
$\rho=\rho(t)$ and $p=p(t)$. If one can solve the first equation with
respect to $t$ as $t=t(\rho)$,
by substituting it into the second equation,  an equation of state (EOS)
follows:
\be
\label{PQE54}
p=p\left(t(\rho)\right)\ .
\ee
Of course, $\rho$ and $p$ could be a sum with the contribution of several kinds of fluids with simple EOS.

We now concentrate on the case that $f(R)$ is given by\cite{NO}
\be
\label{PQE55}
f(R)= - \frac{\alpha}{R^n} + \frac{R}{2\kappa^2} + \beta R^2\ .
\ee
Furthermore, we write $H(t)$ as Eq. (\ref{PQR29}) and assume $h(t)$ is
slowly varying function
of $t$ and neglect the derivatives of $h(t)$ with respect to $t$.
Then one gets (\ref{PQE56}).

First we consider the case that the last term in (\ref{PQE51}) dominates $f(R)\sim \beta R^2$, which may
correspond to the early (inflationary) epoch of the universe. It is not
difficult to
find
\bea
\label{PQE57}
\rho &\sim& -\frac{36 \beta \left(-1 + 2h(t)\right) h(t)^2 }{t^{4n}}\ ,\nn
p & \sim & - \frac{36 \beta \left(-1 + 2h(t)\right) h \left(3h(t) + 1\right)}{t^{4n}}\ .
\eea
If $h$ goes to infinity, which corresponds to deSitter universe, we find
$\rho\sim p \sim h^3$ although,
 from (\ref{PQE56}), $R\sim h^4$. Therefore $\rho$, $p  \ll \beta R^2$ and contribution form the
matter could be neglected. Then the inflation could be generated
only by the contribution from the higher
curvature term.

Second, we consider the case that the second term in (\ref{PQE51})
dominates $f(R)\sim \frac{R}{2\kappa^2}$,
which may correspond to the matter dominated epoch after the inflation. In
this case $\rho$ and $p$ behave as
\bea
\label{PQE58}
\rho &\sim & \frac{12 h(t) + 6h(t)^2}{\kappa^2 t^2}\ ,\nn
p &\sim& - \frac{4h(t) - 6h(t)^2}{\kappa^2 t^2}\ .
\eea
In the matter dominated epoch, we expect $h\sim 2/3$ $\left(a\sim
t^{\frac{2}{3}}\right)$. Hence, one gets
\be
\label{PQE59}
\rho \sim \frac{32}{3\kappa^2 t^2}\ ,\quad p \sim 0\ .
\ee
Therefore in the matter sector, dust with $w=0$ ($p=0$) should dominate, as usually expected.

Finally we consider the case that the first term in (\ref{PQE52}) dominates $f(R)\sim - \alpha/R^n$,
which might describe the acceleration of the present universe. The behavior of of $\rho$ and $p$
is given by
\bea
\label{PQE60}
\rho&\sim& \alpha\left\{6\left(n+1\right)\left(2n + 1\right)h(t) + 6 \left(n-2\right)h(t)^2\right\} \nn
&& \times \left\{-6h(t) + 12 h(t)^2 \right\}^{-n-1}t^{2n} \ ,\nn
p&\sim& \alpha\left\{ -4n\left(n+1\right)\left(2n+1\right) -2 \left(8 n^2 + 5n +3\right)h(t) \right. \nn
&& \left.  - 6 \left(n-2\right)h(t)^2\right\} \nn
&& \times \left\{-6h(t) + 12 h(t)^2 \right\}^{-n-1} t^{2n}\ .
\eea
Thus, the effective EOS parameter $w_l$ is given by
\bea
\label{PQE61}
w_l &\equiv& \frac{p}{\rho} \nn
&\sim& \left\{ -4n\left(n+1\right)\left(2n+1\right) -2 \left(8 n^2 + 5n +3\right)h(t) \right. \nn
&& \left.  - 6 \left(n-2\right)h(t)^2\right\} \left\{6\left(n+1\right)\left(2n + 1\right)h(t) \right. \nn
&& \left. + 6 \left(n-2\right)h(t)^2\right\}^{-1}\ .
\eea
In order that the acceleration of the universe could occur, we find $h>1$.
Let us  now assume that $h(t)\to h_f$ when $t\to \infty$. Then one obtains
\bea
\label{PQE64}
w_l &\to& w_f \nn
&\equiv& \left\{ -4n\left(n+1\right)\left(2n+1\right) -2 \left(8 n^2 + 5n +3\right)h_f \right. \nn
&& \left.  - 6 \left(n-2\right)h_f^2\right\} \left\{6\left(n+1\right)\left(2n + 1\right)h_f \right. \nn
&& \left. + 6 \left(n-2\right)h_f^2\right\}^{-1}\ ,
\eea
and $H(t)\to h_f/t$. Since the matter energy density $\rho_{w_f}$ with the EoS parameter $w_f$ behaves as
\be
\label{PQE65}
\rho_{w_f} \propto a^{-3(1+w_f)} \propto \exp \left( -3(1+w_f)\int dt H(t) \right)\ ,
\ee
the energy density is
\be
\label{PQE66}
\rho_{w_f} \propto t^{-3(1+w_f)h_f}\ .
\ee
Comparing (\ref{PQE66}) with (\ref{PQE60}), we find
\be
\label{PQE67}
2n = -3(1+w_f)h_f\ ,
\ee
which can be confirmed directly from (\ref{PQE64}).

>From the above consideration, we find $\rho$ and $p$ contain mainly
contributions from dust with $w=0$,
$\rho_d(t)$, $p(t)=0$ and ``dark energy'' with $w=w_l$ in (\ref{PQE61}), $\rho_l(t)$, $p_l(t)$. In the expressions
of $\rho(t)$ and $p(t)$ in (\ref{PQE53}), there might be a remaining part:
\be
\label{PQE62}
\rho_R(t) \equiv \rho(t) - \rho_d(t) - \rho_l(t)\ ,\quad
p_R(t) \equiv p(t) - p_l(t)\ ,
\ee
which may help the transition from the matter dominated epoch to the
acceleration epoch. By deleting $t$ in the
expression of (\ref{PQE62}), we obtain the EOS for the remaining part:
\be
\label{PQE63}
p_R=p_R(\rho_R)\ ,
\ee
which may be called the compensating dark energy.
More concretely, according to (\ref{PQE59}), one may have
\be
\label{PQE68}
\rho_d \sim \frac{32}{3\kappa^2 t_0^2}\e^{-3\int_{t_0}^t dt \frac{h(t)}{t}}\ ,
\ee
and according to (\ref{PQE60}),
\bea
\label{PQE69}
\rho_l&\sim & \alpha\left\{6\left(n+1\right)\left(2n + 1\right)h_f + 6 \left(n-2\right)h_f^2\right\} \nn
&& \times \left\{-6h_f + 12 h_f^2 \right\}^{-n-1}t_1^{2n}\e^{-3(1+w_f)\int_{t_1}^t dt \frac{h(t)}{t}}\ .
\eea
In (\ref{PQE69}), we choose $t_1$ to be large enough. When $t\sim t_0$,
 $\rho(t) \sim \rho_d$
and when $t\to \infty$, $\rho(t) \sim \rho_l$. Thus, $\rho_R$ only
dominates after $t=t_1$ but it becomes
smaller in late times. Hence, the role of $\rho_R$ (which perhaps may be
identified partially with dark matter) is only to connect the matter
dominated epoch to the acceleration epoch.

\section{Discussion}

In summary, we developed the general formulation of modified $f(R)$
gravity which may be reconstructed for any given FRW metric.
The resulting action is given in the implicit form (usually, in terms of
some special functions). Nevertheless, its early and late times
asymptotics may be defined, it turns out to have quite simple form,
for instance, as model \cite{NO} with negative and positive powers of
curvature. Several examples predicted by realistic cosmology are constructed.
These specific $f(R)$ theories describe the sequence of cosmological
epochs: matter dominated stage (if necessary even without matter),
transition from decceleration to acceleration and current cosmic speed-up
consistent with three years WMAP data.
Moreover, the study of their newtonian regime indicates that such
models are consistent with Solar System tests. It is not difficult to
extend such a formulation to include consistently also radiation dominated
phase (perhaps, even inflation). Hence, modified $f(R)$ gravity indeed
represents the realistic alternative to General Relativity, being more
consistent in dark epoch.
It is also shown that some implicit version of $f(R)$ gravity may describe
$\Lambda$CDM cosmology without need to introduce the cosmological constant
 and without singularity near $R=0$.

In the alternative approach we also demonstrate that even simple models
like the ones of ref.\cite{NO,CNOT} become cosmologically viable if
compensating dark energy is introduced. It remains to study if such compensating
dark energy or the version of $f(R)$ gravity which mimics matter dominated phase
without matter may serve as dark matter of the universe.

Definitely, more careful study of modified gravity and fitting the above
models against the observational data/various constraints \cite{tegmark}
should be done.
First of all, one should study the linear perturbations in the
matter-dominated epoch. Such a study made for scalar-Gauss-Bonnet gravity
model of ref.\cite{sasaki}
in ref.\cite{koivisto} shows that there is no really strong change if
compare with usual General Relativity. Hence, one may expect that
the same will occur in the present model while careful study should be
made, of course. It is also known \cite{constrain} that Supernovae Ia
constraints are easy
to fit in the class of models under consideration (subject to assumption
that they are treated as usual candles).
Second, the CMBR peak locations are weakly model-dependent. Nevertheless,
it is important to check constraints appearing from the CMBR shift
parameter and baryon oscillations as well as nucleosynthesis bounds which
restrict the amount of dark energy in the current universe.
Third, Solar System constraints (time variation of the effective
gravitational constant, study of PPN parameters) should be considered in
detail. The preliminary expectation is that this may be achieved due to the
fact of freedom of the form of modified gravity (as well as first time
derivative of the gravitational Lagrangian) at some specific
(initial) time.
This will be investigated in detail elsewhere.
Having in mind that new, more precise observational data will be available
soon, one may expect that the question:
is modified gravity suitable as dark energy will be answered in near
future.

\section*{Acknowledgements}

We thank S. Capozziello for helpful discussions.
The investigation by S.N. has been supported in part by the
Ministry of Education, Science, Sports and Culture of Japan under
grant no.18549001
and 21st Century COE Program of Nagoya University
provided by Japan Society for the Promotion of Science (15COEG01),
and that by S.D.O. has been supported in part  by the
project FIS2005-01181
(MEC,Spain), by the project 2005SGR00790 (AGAUR,Catalunya), by LRSS
project N4489.2006.02 and by RFBR grant 06-01-00609
(Russia).

\appendix

\section{}
Let us show for the explicit example of $f(R)$ that our formulation
with auxiliary scalar fixed as time is
equivalent to usual metric formulation.
The starting action of the modified gravity coupled with matter is:
\be
\label{M1}
S = \int d^4 x \sqrt{-g}\left\{ f_0 R^\alpha + {\cal L}_{\rm matter}\right\}\ .
\ee
FRW equation is given by
\bea
\label{M7}
0&=& f_0 \left\{ - \frac{1}{2} \left(6\dot H + 12 H^2\right)^\alpha \right. \nn
&& + 3 \alpha \left(\dot H + H^2\right)\left(6\dot H + 12 H^2\right)^{\alpha - 1} \nn
&& \left. -3\alpha H\partial_t\left\{\left(6\dot H + 12 H^2\right)^{\alpha - 1}\right\}\right\} \nn
&& + \frac{1}{2}\rho_0 a^{-3(1+w)}\ .
\eea
An exact solution of (\ref{M7}) is given by
\bea
\label{M8}
&& a=a_0 t^{h_0} \ ,\quad h_0\equiv \frac{2\alpha}{3(1+w)} \ ,\nn
&& a_0\equiv \left[-\frac{6f_0h_0}{\rho_0}\left(-6h_0 + 12 h_0^2\right)^{\alpha-1} \right. \nn
&& \left. \times
\left\{\left(1-2\alpha\right)\left(1-\alpha\right) - (2-\alpha)h_0\right\}\right]^{-\frac{1}{3(1+w)}}\ .
\eea

Instead of action (\ref{M1}), one now starts with the action (\ref{PQR1}).
Eq.(\ref{M8}) shows
\be
\label{EX1}
H=\frac{h_0}{t}\ \mbox{or}\ g(t)=h_0 \ln t\ .
\ee
Hence
\bea
\label{PQR11b}
0&=&2 \frac{d^2 P(\phi)}{d\phi^2} - 2 g'(\phi) \frac{dP(\phi))}{d\phi} + 4g''(\phi) P(\phi) \nn
&& + \left(1 + w\right) \rho_{0} a_0^{-3(1+w)} \e^{-3(1+w)g(t)} \nn
&=& 2 \frac{d^2 P(\phi)}{d\phi^2} - \frac{2 h_0}{\phi} \frac{dP(\phi))}{d\phi} - \frac{4 h_0}{\phi^2} P(\phi) \nn
&& + \left(1 + w\right) \rho_{0} a_0^{-3(1+w)} \phi^{-2\alpha} \ .
\eea
Here we have used a relation $h_0=(2/3)\left(\alpha/(1+w)\right)$ in (\ref{M8}).
A solution of (\ref{PQR11b}) is given by
\bea
\label{EX2}
P &=& P_0\phi^{-2\alpha + 2}\ ,\nn
P_0 &\equiv& \frac{(1+w)\rho_0 a_0^{-3(1+w)}}{4\left\{(1-\alpha)(1-2\alpha) - h_0 (2 - \alpha)\right\}}\ .
\eea
Then we find
\bea
\label{PQR12b}
Q(\phi)&=&-6 \left(g'(\phi)\right)^2 P(\phi)
 - 6g'(\phi) \frac{dP(\phi)}{d\phi} \nn
&& + \rho_{0} a_0^{-3(1+w)}  \e^{-3(1+w)g(t)} \nn
&=&Q_0 \phi^{-2\alpha} \ ,\nn
Q_0&\equiv & - 6h_0\left(h_0 - 2\alpha + 2\right)P_0 + \rho_0 a_0^{-3(1+w)} \nn
&=& \frac{6(1-\alpha)h_0(2h_0 - 1) P_0}{\alpha}\ .
\eea
In the last line, we used the definition of $P_0$ in (\ref{EX2})
Therefore it follows
\bea
\label{PQR2b}
0&=&P'(\phi)R + Q'(\phi) \nn
&=&2(1-\alpha) \phi^{-2\alpha+1} R - 2\alpha Q_0 \phi^{-2\alpha}\ ,
\eea
which gives
\be
\label{EX3}
\phi^2=\frac{\alpha Q_0}{(1-\alpha)P_0 R} = \frac{12h_0^2 - 6h_0}{R}\ .
\ee
Then the action(\ref{PQR1}) has the following form
\bea
\label{M1b}
S &=& \int d^4 x \sqrt{-g}\left\{ f_0' R^\alpha + {\cal L}_{\rm matter}\right\} \ , \nn
f_0' &\equiv & \left(\frac{\alpha}{1 - \alpha}\right)^{-\alpha + 1}\frac{Q_0^{-\alpha+1} P_0^\alpha}{\alpha} \nn
&=&\frac{\left(12h_0^2 - 6 h_0\right)^{-\alpha + 1} \rho_0 a_0^{-3(1+w)}}
{6\left\{(1-\alpha)(1-2\alpha) - h_0 (2-\alpha)\right\}}\ .
\eea
Comparing $f_0'$  (\ref{M1b}) with (\ref{M8}), one gets
\be
\label{EX4}
f_0=f_0'\ ,
\ee
which shows the action (\ref{M1}) is surely reproduced.
On the other hand, Eq.(\ref{M8}) shows
\be
\label{EX5}
R=\frac{12h_0^2 - 6h_0}{t^2}\ .
\ee
Comparing (\ref{EX5}) with (\ref{EX3}), we find
\be
\label{EX6}
\phi=t\ .
\ee

\end{document}